\documentclass[letterpaper]{article} 
\usepackage{aaai25}  
\usepackage{times}  
\usepackage{helvet}  
\usepackage{courier}  
\usepackage[hyphens]{url}  
\usepackage{graphicx} 
\usepackage{natbib}  
\usepackage{caption} 
\usepackage{algorithm}
\usepackage{algorithmic}
\usepackage[table,xcdraw]{xcolor}
\usepackage{float}
\usepackage{cite}
\usepackage{amsmath,amssymb,amsfonts}
\usepackage{multirow}
\usepackage{utfsym}
\usepackage{makecell}
\usepackage{textcomp}
\usepackage{newfloat}
\usepackage{listings}
\DeclareCaptionStyle{ruled}{labelfont=normalfont,labelsep=colon,strut=off} 
\floatstyle{ruled}
\newfloat{listing}{tb}{lst}{}
\floatname{listing}{Listing}
\title{Motion Artifact Removal in Pixel-Frequency Domain\\via Alternate Masks and Diffusion Model}
\author{
    Jiahua Xu\textsuperscript{\rm 1}\equalcontrib, 
    Dawei Zhou\textsuperscript{\rm 1}\equalcontrib, 
    Lei Hu\textsuperscript{\rm 2}\equalcontrib,
    Jianfeng Guo\textsuperscript{\rm 3},
    Feng Yang\textsuperscript{\rm 3}, \\
    Zaiyi Liu\textsuperscript{\rm 2}\footnote{Corresponding authors.}, 
    Nannan Wang\textsuperscript{\rm 1}\footnotemark[2],
    Xinbo Gao\textsuperscript{\rm 4}
}
\affiliations{
    \textsuperscript{\rm 1}State Key Laboratory of Integrated Services Networks, Xidian University, Xi’an, China\\
    \textsuperscript{\rm 2}Department of Radiology, Guangdong Provincial People’s Hospital, Southern Medical University, Guangzhou, China\\
    \textsuperscript{\rm 3}Department of Radiology, Xiangyang No. 1 People’s Hospital, Hubei University of Medicine, Xiangyang, China\\
    \textsuperscript{\rm 4}Chongqing Key Laboratory of Image Cognition, Chongqing University of Posts and Telecommunications, Chongqing, China\\
    jhxu.xidian@gmail.com, dwzhou.xidian@gmail.com, hulei@gdph.org.cn, jianfengguo@hbmu.edu.cn,\\
    haitang76@163.com, liuzaiyi@gdph.org.cn, nnwang@xidian.edu.cn, gaoxb@cqupt.edu.cn
}

\usepackage{bibentry}

\begin{document}

\maketitle

\begin{abstract}
Motion artifacts present in magnetic resonance imaging (MRI) can seriously interfere with clinical diagnosis. Removing motion artifacts is a straightforward solution and has been extensively studied. However, paired data are still heavily relied on in recent works and the perturbations in \textit{k}-space (frequency domain) are not well considered, which limits their applications in the clinical field. To address these issues, we propose a novel unsupervised purification method which leverages pixel-frequency information of noisy MRI images to guide a pre-trained diffusion model to recover clean MRI images. Specifically, considering that motion artifacts are mainly concentrated in high-frequency components in \textit{k}-space, we utilize the low-frequency components as the guide to ensure correct tissue textures. Additionally, given that high-frequency and pixel information are helpful for recovering shape and detail textures, we design alternate complementary masks to simultaneously destroy the artifact structure and exploit useful information. Quantitative experiments are performed on datasets from different tissues and show that our method achieves superior performance on several metrics. Qualitative evaluations with radiologists also show that our method provides better clinical feedback. Our code is available at \url{https://github.com/medcx/PFAD}.
\end{abstract}

%

\section{Introduction}\label{sec:introduction}
Magnetic resonance imaging (MRI) is a 
non-invasive medical imaging technique and has been widely used in clinical diagnostics. However, it usually requires extended acquisition time \cite{zaitsev2015motion} and some patients (\textit{e.g.}, elderly or those with conditions such as claustrophobia and epilepsy \cite{chen1990linear}) may experience motions due
to discomfort during the long scanning duration. These motions may destroy the \textit{k}-space (aka frequency domain) signal distribution of MRI, leading to motion artifacts and thus seriously interfering with clinical diagnosis and treatment. Therefore, we need to explore how to effectively alleviate the affects of motion artifacts. 

\begin{figure}[t]
    \centering
    \includegraphics[width=3.05in]{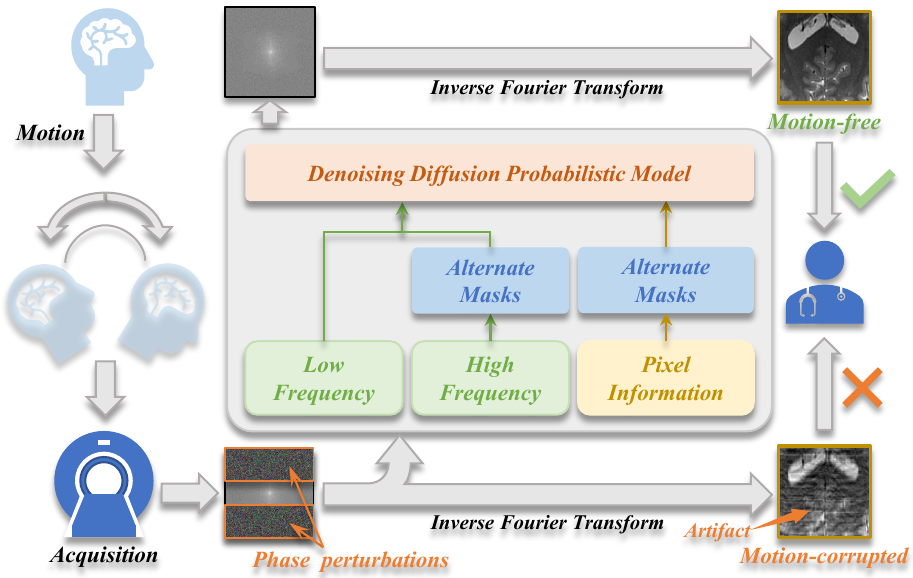}
    \caption{Schematic of our research problem and solution. During the MRI imaging process, motion artifacts are produced due to the patient's motion. Removing artifacts via a diffusion model and alternate masks in pixel-frequency domain can help radiologists make better diagnosis.}
    \label{motivation_pdf}
\end{figure}

Several approaches have been proposed to solve this problem \cite{hu2024protecting, hu2024development}. They can be classified into prospective motion artifact prevention and retrospective motion artifact removal. For the former, researchers aim to preemptively avoid motion artifacts by tracking the motion \cite{schulz2012embedded, zaitsev2006magnetic}, such as modifying imaging sequences and parameters \cite{cruz2016accelerated, vasanawala2010navigated, cruz2017highly}. But this strategy often entails additional equipment or prolonged scanning times, resulting in higher costs. 
For the latter, researchers use compressed sensing to iteratively correct motion artifacts \cite{vasanawala2010improved, yang2013sparse, kustner2017self, hansen2012retrospective}. However, these approaches typically require raw \textit{k}-space data which poses a data barrier to effectively building models. Additionally, traditional approaches usually require paired data for training \cite{isola2017image,tamada2020motion}, but paired data is hard and costly to obtain due to changed tissue positioning at rescanning.

Recently, denoising diffusion probabilistic model (DDPM) have demonstrated outstanding performance in the field of image generation \cite{song2020denoising, ho2022classifier}. It has been verified that incorporating guidance such as images \cite{rombach2022high, saharia2022palette} or texts \cite{nichol2021glide, ramesh2022hierarchical, saharia2022photorealistic, si2023freeu, zhang2023adding} during the inference process of DDPM can facilitate the generation of desired images tailored to specific requirements. However, despite the excellent ability of diffusion-based approaches to purify or reconstruct natural images, the performance in motion artifacts removal of MRI images is still limited. This is because motion artifacts exist in \emph{k}-space whereas only pixel-domain information is of interest during their purification process. 

To address above issues, as shown in Figure~\ref{motivation_pdf}, we propose a method called PFAD, which removes motion artifacts in Pixel-Frequency domain via Alternate masks and Diffusion model. 
Specifically, we first pre-train a diffusion model using unpaired clean images to \textit{fit the distribution of MRI data without motion artifacts}. 
Secondly, we extract the low-frequency information of the input noisy images to \textit{guide the generation of clean images with correct tissue textures}. Then, given that high-frequency and pixel information are helpful for recovering shape and detail textures, we construct alternate masks for the high-frequency/pixel components and generated image in the inference process to \textit{destroy the artifact structure and preserve partial useful information for recovering}. The masks are flipped at each reverse step to alternatively cover all image areas. Following this, we introduce a parameter that changes with diffusion time to further disrupt the distribution of artifacts. Experimental results show that this parameter plays a crucial role in improving the quality of recovered images. 
Finally, we design a hyperparameter to balance the recovered signals from frequency and pixel domains to obtain best clean images.

The main contributions are summarized as follows:
\begin{itemize}
  \item An unsupervised method is proposed for motion artifact removal, which jointly leverages the frequency and pixel information of MRI images to purify noisy images without requesting paired clean data.
  \item We construct alternate complementary masks for high-frequency and pixel components in the reverse process to effectively destroy the artifact structure while retaining useful information for recovering clean signals.
  \item We design two parameters to dynamically adjust the ratio of guidance from the noisy and generated images, and trade off the recovered signals from the pixel and frequency domains, respectively, to simultaneously promote artifact removal and image quality.
  \item Quantitative and qualitative experiments are conducted on multiple datasets with different tissues, which demonstrates the of our method in metrics and clinical assessment. Ablation studies are performed to validate the effectiveness of each module in our method.
\end{itemize}
  
\section{Related Work}
\subsection{MRI Motion Artifacts}
The previous work \cite{budde2014ultra} noted that during the encoding in MRI images, patients' movements can cause errors in \emph{k}-space along specific phase encoding lines, resulting in blurring or ghosting. Therefore, refer to \cite{oh2023annealed}, we adopt a lenient assumption that the motion artifacts in MRI images predominantly manifest in the high-frequency region along the phase encoding lines. It is due to the fact that \emph{k}-space data acquisition usually starts from the centre region, during which the patient usually does not not move spontaneously. It is only after a period of time does the patient move, resulting in perturbation of phase information in the high frequency region. Consequently, our focus lies on removing artifacts in the high-frequency region, and set it to be greater than $\pi$/10 \cite{oh2023annealed}.

\subsection{Motion Artifact Removal}
Motion artifacts can significantly disrupt clinical diagnosis and treatment. With the rapid development of deep learning, self-supervised convolutional neural networks have gradually been applied to motion artifact removal in MRI images. 

Researchers used GAN-based approaches, such as Pix2pix \cite{isola2017image}, CycleGAN \cite{zhu2017unpaired}, CycleMedGAN-V2.0 \cite{armanious2020unsupervised} and stochastic degradation block \cite{chung2021simultaneous} to remove artifacts. 
However, in reality, the distribution of motion artifacts is variable due to changes in motion forms, so these methods are difficult to cope with complex clinical scenario. Thus, researchers tried to recover images using diffusion models such as DR2 \cite{wang2023dr2} and GDP \cite{fei2023generative}. 

The above approaches are difficult to apply in the clinical diagnosis due to the neglect of \emph{k}-space phase perturbations. Therefore, researchers proposed UDDN \cite{wu2023unsupervised} to remove motion artifacts in frequency domain. Meanwhile, \cite{oh2023annealed} used the score-based diffusion model to perform the recovery using measured $k$-space values. Noting that most previous works are based on GANs and remove artifacts only in pixel domain, we propose a novel method based on DDPM in pixel-frequency domain.

\section{Methodology}
\subsection{Preliminary}
\subsubsection{Notice}
We use lower-case letters $x$ to denote the image. Moreover, we use $t$ to represent the timestep that the diffusion model is going through and $T$ to represent the total number of timesteps, with $t \in \{1,\ldots, T\}$. We use $\mathcal{N}(\cdot;\mu, \sigma^2I)$ to denote a variable that obeys a Gaussian distribution with mean $\mu$ and variance matrix $\sigma^2$. We denote a diffusion model by $\mathcal{D}_\theta$, where $\theta$ is its model parameters. We use $\mathcal{F}$ to represent the Fourier transform, $\mathcal{F}^{-1}$ to represent the inverse Fourier transform, and $\Phi_l$ and $\Phi_h$ to represent a pair of complementary ideal low-pass filter and high-pass filter with the cutoff frequency of $\pi$/10. We use $|\cdot|$ to represent the modulo value operation and $\odot$ to represent the Hadamard product.

\subsubsection{Diffusion Model}
The diffusion model is a kind of generative model \cite{ho2020denoising, dhariwal2021diffusion, song2020denoising, saharia2022palette, rombach2022high}. Specifically, given an image $x_0$, the forward process gradually adds noise to obtain $x_1$, $x_2, \ldots, x_T$. This is represented as follows,
\begin{equation}
\label{eq:2}
q(x_t|x_{t-1})=\mathcal{N}\left(x_t;\sqrt{1-\beta_t}x_{t-1}, \beta_tI\right),
\end{equation}
where $\beta_t \in \{\beta_1, \ldots, \beta_T\}$ represents a variable that grows linearly with the timesteps. $I$ is the identity matrix.

At the same time, we can also get the reverse process,
\begin{equation}
\label{eq:8}
p_\theta(x_{t-1}|x_t)=\mathcal{N}\left(x_{t-1};\mu_\theta(x_t,t),\sigma_t^2I\right),
\end{equation}
where $\mu_\theta$ denotes a trainable neural network, $\sigma^2_t$ is always set as constants, and the iteration starts from $x_T\sim\mathcal{N}(0, I)$.

To train the neural network, we can obtain the optimization objective as follows,
\begin{equation}
\label{eq:9}
\mathcal{L}_{diff}=\mathbb{E}_{x_0, \epsilon \sim \mathcal{N}(0,I)}\left[\Vert \epsilon-\epsilon_\theta(\sqrt{\bar{\alpha}_t}x_0+\sqrt{1-\bar{\alpha}_t}\epsilon,t) \Vert^2\right],
\end{equation}
where $\epsilon_\theta$ denotes the noise prediction network based on U-Net \cite{ronneberger2015u} with parameter $\theta$ and $\bar{\alpha}_t=\prod_{i=1}^{t}(1-\beta_i)$.

\subsection{Motion Artifact Removal with PFAD}
\begin{figure*}[!t]
    \centering
    \includegraphics[width=6.05in]{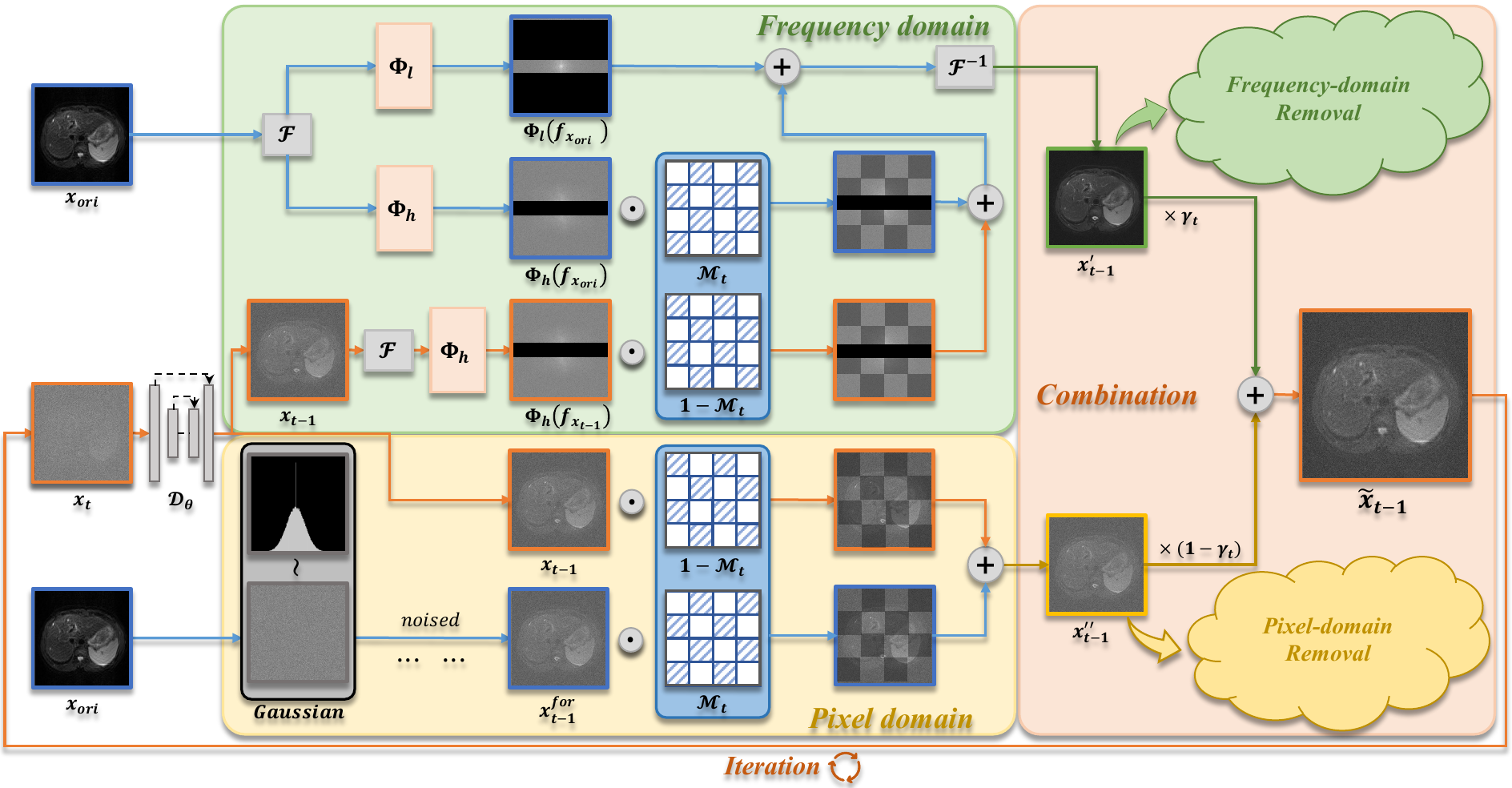}
    \caption{The framework of PFAD. The green area is the artifact removal process in frequency domain. This process is guided by the low-frequency information and part of the high-frequency information of the motion-corrupted image, and uses the diffusion model to generate another part of clean high-frequency information. The yellow area is the pixel domain artifact removal process, which is guided by the partial image information of the forward process. After combining the both domains results, the clean image is finally generated iteratively.}
    \label{method}
\end{figure*}

{\label{framework}}
In this section we will introduce the principle and process of PFAD. The framework is shown in Figure~\ref{method}. Firstly, we introduce how to use alternate complementary masks in frequency domain and pixel domain, respectively. Secondly, we introduce how to create alternate complementary masks. Finally, we design a hyperparameter to balance the weights of the frequency domain and pixel domain. The details of the overall process are presented in Algorithm~\ref{alg}.

\subsubsection{Frequency Domain Removal}\label{freq_domain}
We first focus on motion artifact removal in frequency domain as shown in Figure~\ref{method}. 

Firstly, we extract the low-frequency information from the $x_{ori}$ for subsequent information reorganization, 
\begin{equation}
\label{eq:13}
\Phi_l(f_{x^\prime_{t-1}})=\Phi_l(f_{x_{ori}}),
\end{equation}
where $f$ denotes the frequency information of the image, $x_{ori}$ denotes the motion-corrupted image and $x^\prime$ is the reorganised image of frequency domain.

After that, we apply the alternate complementary masks in the high-frequency region,
\begin{equation}
\label{eq:114}
\Phi_h(f_{x^\prime_{t-1}})=\Phi_h(f_{x_{ori}})\odot \mathcal{M}_t+\Phi_h(f_{x_{t-1}})\odot\left(1-\mathcal{M}_t\right),
\end{equation}
where ${\mathcal{M}}_t$ is the alternate complementary masks, which will be introduced later.

By combining the reorganised high-frequency information with the original low-frequency information, we can obtain the reorganised image, as follows,
\begin{equation}
\label{eq:14}
x_{t-1}^\prime=\Big|\mathcal{F}^{-1}\left(\Phi_h(f_{x^\prime_{t-1}})+\Phi_l(f_{x^\prime_{t-1}})\right)\Big|.
\end{equation}

\subsubsection{Pixel Domain Removal}\label{image_domain} 
Focusing only on the recovery of high-frequency details while ignoring low-frequency information will result in an overall lack of naturalness in the image. In order to improve the naturalness of the image and further reduce artifacts, we also perform motion artifact removal in the pixel domain. At the same time, pixel domain removal can also deal with the slight phase loss caused by the modulo value operation in frequency domain removal.

As shown in Figure~\ref{method}, unlike motion artifact removal in frequency domain, we replace the guidance information $x_{ori}$ with $x^{for}_{t-1}$ of 
the forward process. Employing this method in pixel domain allows for better utilization of the inherent generation capability of the diffusion model, resulting in clearer images \cite{lugmayr2022repaint}. The expression is as follows,
\begin{equation}
\label{eq:15}
x_{t-1}^{\prime\prime}=x_{t-1}^{for}\odot\mathcal{M}_t+x_{t-1}\odot(1-\mathcal{M}_t),
\end{equation}
where $x^{\prime\prime}$ is the reorganised image of pixel domain and $x^{for}$ denotes the image of the forward process. 
\begin{algorithm}[t]
    \footnotesize
    \caption{PFAD for motion artifact removal}
    \label{alg}
    \renewcommand{\algorithmicrequire}{\textbf{Input:}}
    \renewcommand{\algorithmicensure}{\textbf{Output:}}
    \renewcommand{\algorithmiccomment}[1]{\hfill $\triangleright$ #1}
    \begin{algorithmic}[1]
        \renewcommand{\baselinestretch}{1} 
        \selectfont
        \REQUIRE  $x_{ori}$, $x_T \sim \mathcal{N}\left(0,I\right)$, $\Phi_h$, $\Phi_l$, $a$, $D_\theta$, $\{\bar{\alpha}_i\}^T_1$, $\{m_j\}^T_1$.
        \ENSURE The motion-free image $\widetilde{x}_0$.
        \FOR{$i=T$ to 1}
            \STATE $x_{i-1} \gets D_\theta(x_i)$ 
            \STATE $\omega_i \gets 1-\sqrt{\bar{\alpha}_i}$
            \STATE $\mathcal{M}_{i} \gets \omega_i\cdot m_{i}$
            \STATE $\Phi_l(f_{x^\prime_{i-1}}) \gets \Phi_l(f_{x_{ori}})$
            \STATE $\Phi_h(f_{x^\prime_{i-1}}) \gets \Phi_h(f_{x_{ori}})\odot \mathcal{M}_i+\Phi_h(f_{x_{i-1}})\odot\left(1-\mathcal{M}_i\right)$
            \STATE $x_{i-1}^\prime \gets |\mathcal{F}^{-1}(\Phi_h(f_{x^\prime_{i-1}})+\Phi_l(f_{x^\prime_{i-1}}))|$  \COMMENT{Frequency domain}
            \STATE $\epsilon \sim \mathcal{N}(0,I)$
            \STATE $x_{i-1}^{for} \gets \sqrt{\bar{\alpha}_i}\cdot x_{ori}+\sqrt{1-\bar{\alpha}_i}\cdot \epsilon$ 
            \STATE $x_{i-1}^{\prime\prime} \gets x_{i-1}^{for}\odot\mathcal{M}_i+x_{i-1}\odot(1-\mathcal{M}_i)$ \COMMENT{Pixel domain}
            \STATE $\gamma_i \gets -a\cdot e^{-\frac{i}{T}}+1$
            \STATE $\widetilde{x}_{i-1} \gets \gamma_i\cdot x_{i-1}^\prime+(1-\gamma_i)\cdot x_{i-1}^{\prime\prime}$ \COMMENT{Dual domain balance}
            \STATE $x_{i-1} \gets \widetilde{x}_{i-1}$ 
        \ENDFOR
        \RETURN $\widetilde{x}_0$
        \renewcommand{\baselinestretch}{1} 
        \selectfont
    \end{algorithmic}
\end{algorithm}

\subsubsection{Alternate Complementary Masks}\label{mask}

\begin{figure}[!t]
    \centering
    \includegraphics[width=3.0in]{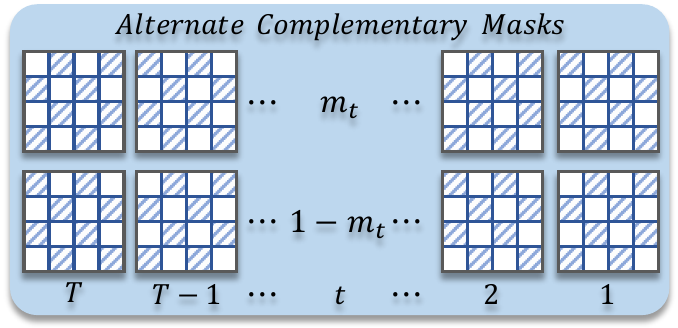}
    \caption{Alternate complementary masks. In each iteration, the complementary masks are alternately transformed to ensure artifact removal for the entire image.}
    \label{mask_pdf}
\end{figure}

Using masks can destroy the artifact distribution while retaining part of the useful information. In previous research, the mask used in diffusion models was fixed  \cite{lugmayr2022repaint}. However, in our method, masks are flipped at each reverse step to alternatively cover all image areas throughout the process, as shown in Figure~\ref{mask_pdf}, which is expressed as follows,
\begin{equation}
\label{eq:115}
m_{t-1}=1-m_{t}.
\end{equation}

During the reverse process, although the alternate complementary masks destroy artifact distribution, the motion artifact information retained by the mask still seriously affect the diffusion model in later stages. Therefore, we introduce a variable to fine-tune the weighting of the masks, aiming to alleviate this influence.
\begin{equation}
\label{eq:116}
\mathcal{M}_{t}=\omega_t\cdot m_{t},
\end{equation}
where $\omega_t=1-\sqrt{\bar{\alpha}_t}$ which is an intrinsic variable of DDPM.

In (\ref{eq:116}), as time $t$ gradually decreases from $T$, $\omega_t$ converges from 1 to 0. This shows that in the initial stage of the reverse process, the model gradually generates the correct shape and textures by useful information. Additionally, as the reorganised image still contains a amount of Gaussian noise at this stage, the influence of motion artifacts can be disregarded. In the later stages, motion artifacts can seriously interfere with the diffusion model. Considering that the generated image already has correct shape and textures at this stage, the model is no longer completely dependent on the guidance of the useful information, so we reduce the value of $\omega_t$ to mitigate the interference of motion artifacts.

\subsubsection{Dual Domain Balance}\label{balance}
In order to remove motion artifacts while ensuring visual clarity and naturalness of the images, we introduce a parameter $\gamma_t$ that varies with the diffusion step $t$,
\begin{equation}
\label{eq:16}
\gamma_t=-a\cdot e^{-\frac{t}{T}}+1,
\end{equation}
where $a$ is a tunable hyperparameter ranging from 0 to 1. $\gamma_t$ is approximately equal to 1 when $t=T$. As $t$ gradually decreases, $\gamma_t$ gradually approaches $1-a$.

We use $\gamma_t$ to balance the recovered images between the frequency domain and pixel domain, as follows:
\begin{equation}
\label{eq:17}
\widetilde{x}_{t-1}=\gamma_t\cdot x_{t-1}^\prime+(1-\gamma_t)\cdot x_{t-1}^{\prime\prime},
\end{equation}
where $\widetilde{x}$ is the final reorganised image.

According to the definition of $\gamma_t$, in the initial stages of the reverse process, we focus more on recovering information in frequency domain to make the reorganised image contain more tissue textures. As $t$ gradually decreases, we gradually increase the proportion of information from the pixel domain, aiming to achieve a dual objective of balancing artifact removal and visual clarity.

\begin{figure*}[t]
    \centering
    \includegraphics[width= 6.7in]{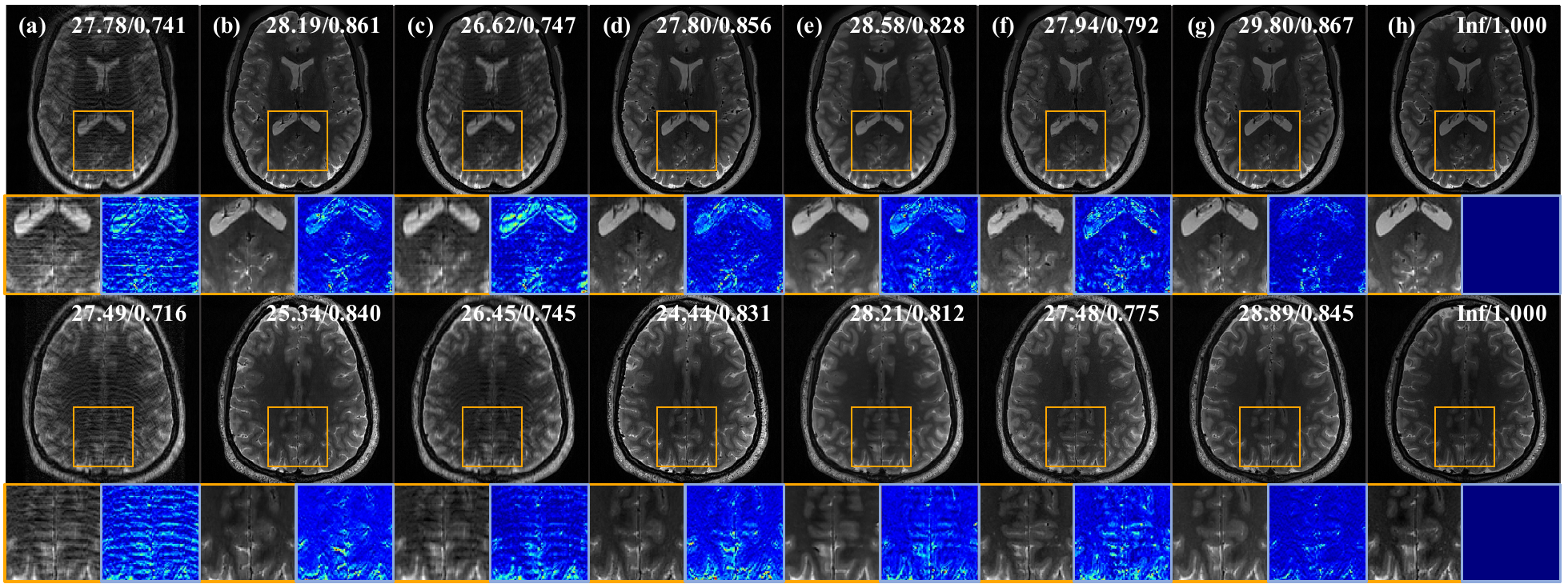}
    \caption{Comparison of results on simulated brain images: (a) Artifact images, (b) Pix2pix, (c) CycleGAN, (d) UDDN, (e) DR2, (f) GDP, (g) PFAD (ours), and (h) the ground-truth. PSNR and SSIM values of each image are shown in the corner of images. The yellow box indicates the zoomed-in visualization area, and the blue box represents the difference heatmap compared to the ground truth. The color ranges from blue to red, indicating differences from small to large, with deeper colors representing smaller differences.}
    \label{show_brain}
\end{figure*}

\begin{table*}[!t]
\begin{small}
\centering
\setlength{\tabcolsep}{8pt}
\begin{tabular}{c|c|ccccccc}
\hline
Dataset & Methods & PSNR $\uparrow$ & SSIM $\uparrow$ & Lpips $\downarrow$ & FSIM $\uparrow$ & VIF $\uparrow$ & VSI $\uparrow$ & GMSD $\downarrow$\\
\hline
\hline
\multirow{7}{*}{\centering \parbox{1.7cm}{\centering Brain \\Random}} & Input & 25.97$^{**}$ & 0.685$^{**}$ & 0.349$^{**}$ & 0.812$^{**}$ & 0.295$^{**}$ & 0.945$^{**}$ & 0.114$^{**}$\\

& Pix2pix & \underline{26.76}$^{**}$ & \textbf{0.841} & \underline{0.151}$^{**}$ & \underline{0.882}$^{**}$ & \underline{0.392}$^{**}$ & \underline{0.966}$^{**}$ & \underline{0.065}$^{**}$\\

& CycleGAN & 25.94$^{**}$ & 0.719$^{**}$ & 0.277$^{**}$ & 0.846$^{**}$ & 0.338$^{**}$ & 0.955$^{**}$ & 0.091$^{**}$\\

& UDDN & 26.07$^{**}$ & 0.823$^{**}$ & 0.163$^{**}$ & 0.879$^{**}$ & 0.372$^{**}$ & 0.965$^{**}$ & 0.065$^{**}$\\

& DR2 & 26.60$^{**}$ & 0.780$^{**}$ & 0.174$^{**}$ & 0.859$^{**}$ & 0.372$^{**}$ & 0.960$^{**}$ & 0.083$^{**}$\\

& GDP & 26.08$^{**}$ & 0.739$^{**}$ & 0.193$^{**}$ & 0.847$^{**}$ & 0.313$^{**}$ & 0.957$^{**}$ & 0.092$^{**}$\\

& \cellcolor[HTML]{E8E5E5}PFAD (ours) & \cellcolor[HTML]{E8E5E5}\textbf{27.60} & \cellcolor[HTML]{E8E5E5}\underline{0.829} & \cellcolor[HTML]{E8E5E5}\textbf{0.147} & \cellcolor[HTML]{E8E5E5}\textbf{0.896} & \cellcolor[HTML]{E8E5E5}\textbf{0.426} & \cellcolor[HTML]{E8E5E5}\textbf{0.970} & \cellcolor[HTML]{E8E5E5} \textbf{0.061}\\
\hline

\multirow{7}{*}{\centering \parbox{1.7cm}{\centering Knee \\Random}} & Input & 29.05$^{**}$ & 0.660$^{**}$ & 0.340$^{**}$ & 0.846$^{**}$ & 0.266$^{**}$ & 0.965$^{**}$ & 0.110$^{**}$\\

& Pix2pix & \underline{29.37}$^{**}$ & \underline{0.705}$^{*}$ & {0.279}$^{**}$ & \underline{0.861}$^{**}$ & 0.296$^{**}$ & \underline{0.970}$^{**}$ & \underline{0.085}$^{**}$\\

& CycleGAN & 27.34$^{**}$ & 0.663$^{**}$ & 0.289$^{**}$ & 0.849$^{**}$ & 0.276$^{**}$ & 0.967$^{**}$ & 0.095$^{**}$\\

& UDDN & 28.50$^{**}$ & 0.653$^{**}$ & \underline{0.270}$^{**}$ & 0.852$^{**}$ & {0.274}$^{**}$ & 0.967$^{**}$ & 0.089$^{**}$\\

& DR2 & 29.23$^{**}$ & 0.686$^{**}$ & 0.333$^{**}$ & 0.835$^{**}$ & \underline{0.300}$^{**}$ & 0.964$^{**}$ & 0.101$^{**}$\\

& GDP & 28.08$^{**}$ & 0.599$^{**}$ & 0.285$^{**}$ & 0.840$^{**}$ & 0.213$^{**}$ & 0.966$^{**}$ & 0.104$^{**}$\\

& \cellcolor[HTML]{E8E5E5} PFAD (ours) & \cellcolor[HTML]{E8E5E5}\textbf{29.64} & \cellcolor[HTML]{E8E5E5}\textbf{0.712} & \cellcolor[HTML]{E8E5E5}\textbf{0.220} & \cellcolor[HTML]{E8E5E5}\textbf{0.869} & \cellcolor[HTML]{E8E5E5}\textbf{0.314} & \cellcolor[HTML]{E8E5E5}\textbf{0.972} & \cellcolor[HTML]{E8E5E5}\textbf{0.080}\\
\hline

\multirow{7}{*}{\centering \parbox{1.7cm}{\centering Abdominal \\Respiratory}} & Input & 35.37$^{**}$ & 0.816$^{**}$ & 0.234$^{**}$ & 0.944$^{**}$ & 0.441$^{**}$ & 0.986$^{**}$ & 0.064$^{**}$\\

& Pix2pix & \underline{36.14}$^{**}$ & \underline{0.902}$^{**}$ & 0.147$^{**}$ & \underline{0.955}$^{**}$ & \underline{0.481}$^{**}$ & \underline{0.988}$^{**}$ & \underline{0.058}$^{**}$\\

& CycleGAN & 34.68$^{**}$ & 0.902$^{**}$ & 0.136$^{**}$ & 0.949$^{**}$ & 0.468$^{**}$ & 0.986$^{**}$ & 0.067$^{**}$\\

& UDDN & 35.14$^{**}$ & 0.896$^{**}$ & \underline{0.117}$^{**}$ & 0.951$^{**}$ & 0.473$^{**}$ & 0.986$^{**}$ & 0.066$^{**}$\\

& DR2 & 34.37$^{**}$ & 0.794$^{**}$ & 0.168$^{**}$ & 0.933$^{**}$ & 0.389$^{**}$ & 0.983$^{**}$ & 0.076$^{**}$\\

& GDP & 33.41$^{**}$ & 0.760$^{**}$ & 0.220$^{**}$ & 0.919$^{**}$ & 0.320$^{**}$ & 0.981$^{**}$ & 0.083$^{**}$\\

& \cellcolor[HTML]{E8E5E5} PFAD (ours) & \cellcolor[HTML]{E8E5E5}\textbf{37.24} & \cellcolor[HTML]{E8E5E5}\textbf{0.918} & \cellcolor[HTML]{E8E5E5}\textbf{0.103} & \cellcolor[HTML]{E8E5E5}\textbf{0.962} & \cellcolor[HTML]{E8E5E5}\textbf{0.544} & \cellcolor[HTML]{E8E5E5}\textbf{0.990} & \cellcolor[HTML]{E8E5E5}\textbf{0.051}\\
\hline
\end{tabular}

\caption{Quantitative comparison of different methods on simulated images. The values in the table represent the average values for the entire test dataset. \textbf{Bold} represents the best and \underline{underlined} represents the second best. All results underwent Mann-Whitney U test, $(^*)$ indicates $p \leq 0.05$, and $(^{**})$ indicates $p \leq 0.01$. } 
\label{table:2}
\end{small}
\end{table*}

\begin{figure*}[t]
    \centering
    \includegraphics[width= 6.7in]{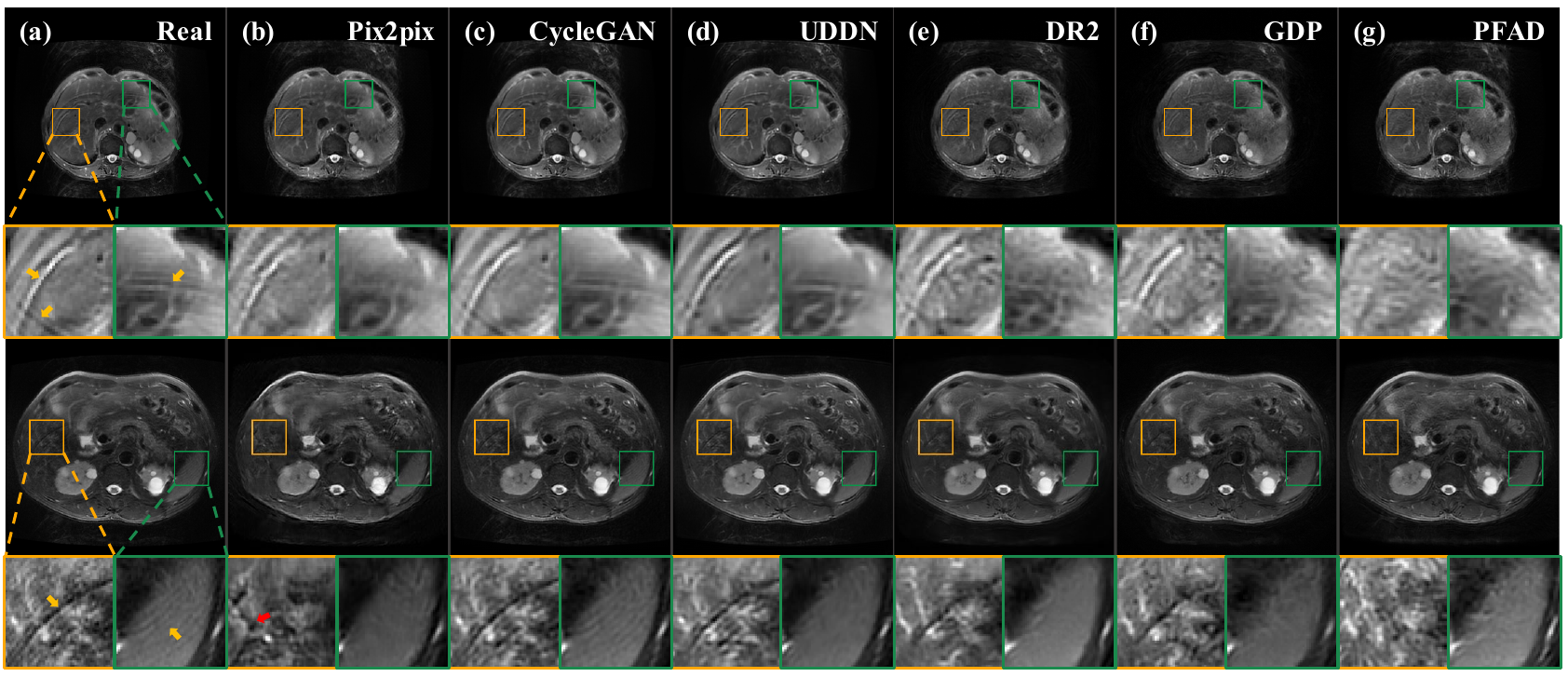}
    \caption{Comparison of results on real images: (a) Images with real motion artifacts, (b) Pix2pix, (c) CycleGAN, (d) UDDN, (e) DR2, (f) GDP, (g) PFAD (ours). The first and third rows show the artifact images in different regions of the abdominal cavity, respectively. And the second and fourth rows show the zoomed-in artifact regions for the first and third rows, respectively. Yellow arrow indicates the location of the artifact, while red arrow points to the area with severe texture deformation.}
    \label{clinical}
\end{figure*}

\section{Implementation Details}
\subsection{Experimental Datasets}
We utilize two public datasets and one private dataset for the evaluation and validation of PFAD. The first dataset is the Human Connectome Project (HCP) data \cite{van2012human}, which is a dataset containing MRI data of the human brain, and the second dataset is the knee MRI data from fastMRI \cite{zbontar2018fastmri}.
The third dataset is the T2-weighted abdominal images from Department of Radiology, Xiangyang No. 1 People’s Hospital, Hubei University of Medicine. The institutional review board approved this study (XYYYE20240082). The more information about above datasets and the simulation of motion artifacts are shown in supplementary material A.1 and A.2.

\subsection{Comparison Approaches}
We select three classes of approaches for comparison. The first is supervised approach, Pix2pix \cite{isola2017image}, which uses paired images for training. The second is unsupervised approaches, CycleGAN \cite{zhu2017unpaired} and UDDN \cite{wu2023unsupervised}, we only use simulated motion-corrupted images for training, as training with real images often leads to unstable training \cite{oh2023annealed}. The third is unsupervised approaches based on diffusion model, DR2 \cite{wang2023dr2} and GDP \cite{fei2023generative}, both are novel approaches for image recovery. The training parameters are set according to the original paper, respectively. For more information on training and evaluation settings, including radiologist evaluation, see supplementary material A.3.

\section{Experimental Results}

\subsection{Quantitative Evaluation of Simulated Images}
Figure~\ref{show_brain} shows the results of various methods for artifact removal on the brain simulated images. The input image contains simulated random motion artifacts (Figure~\ref{show_brain}(a)).

Pix2pix introduces significant deformation in some areas, and its unstable performance leads to a significant decrease in PSNR values for some images as shown in Figure~\ref{show_brain}(b). CycleGAN fails to effectively remove artifacts and results in blurring (Figure~\ref{show_brain}(c)). UDDN fails to maintain the correct tissue textures (Figure~\ref{show_brain}(d)). DR2 tends to mistakenly remove texture details of the brain images, leading to overly smooth images (Figure~\ref{show_brain}(e)). GDP misinterprets some artifacts as brain structures, resulting in images that contain many artifact-like tissue textures (Figure~\ref{show_brain}(e)). In contrast, our method effectively removes artifacts while maintaining correct tissue textures (Figure~\ref{show_brain}(g)). Additionally, our method obtains better quantitative metric values.

We also remove artifacts on knee MRI images. As shown in Figure~\ref{show_knee} in supplementary material B.1, Figure~\ref{show_knee}(a) shows the knee images with simulated motion artifacts. Pix2pix introduces weak ripple-like artifacts (Figure~\ref{show_knee}(b)). Furthermore, CycleGAN significantly changes the contrast of the images (Figure~\ref{show_knee}(c)), and UDDN fails to remove artifacts (Figure~\ref{show_knee}(d)). DR2 and GDP still have corresponding issues: DR2 makes images overly smooth (Figure~\ref{show_knee}(e)), and GDP generates images with artifact-like tissue textures (Figure~\ref{show_knee}(f)). In contrast, our method successfully corrects the artifacts and obtains better quantitative metric values as shown in Figure~\ref{show_knee}(g).

Next, we attempt to remove respiratory motion artifacts on abdominal simulated images. In supplementary material B.1, Figure~\ref{show_abdominal} displays the experimental results with various methods. Pix2pix still exhibits residual artifacts, and the recovered images are overly blurry (Figure~\ref{show_abdominal}(b)). CycleGAN fails to remove artifacts (Figure~\ref{show_abdominal}(c)), leading to degradation in quantitative metrics. Although UDDN induces noticeable texture changes (Figure~\ref{show_abdominal}(d)). DR2 effectively removes respiratory motion artifacts, but some tissue textures are also changed (Figure~\ref{show_abdominal}(e)). In addition, GDP induces significant texture changes in recovered images, resulting in deterioration of the quantitative metrics values (Figure~\ref{show_abdominal}(f)). In contrast, our method not only effectively deals with respiratory motion artifacts, but also successfully reconstructs anatomical structures and details as shown in Figure~\ref{show_abdominal}(g). Our method also obtains higher quantitative metric values.

Table~\ref{table:2} shows the average values of the quantitative metrics. In the knee and abdominal simulated images, our method achieves the highest values across all evaluation metrics. In the brain simulated images, our method is optimal on most metrics, but is lower than Pix2pix on SSIM. This is because Pix2pix is a supervised approach with paired data. However, as shown in Figure~\ref{show_brain}, Figure~\ref{show_knee} and Figure~\ref{show_abdominal}, we find that the performance of Pix2pix is unstable, which results in blurry images and still contains artifacts. Although some metrics of our method are lower than Pix2pix, our method can better remove motion artifacts and recover fine details when processing simulated images.

\subsection{Qualitative Evaluation of Real Clinical Images}

\begin{table}[t]
\begin{small}
\setlength{\tabcolsep}{2.4pt}
\centering
   \begin{tabular}
    {c|ccccc}
    \hline
    Methods & Artifacts & Blurring & Anatomical & Diseased & Overall\\
    \hline
    Pix2pix & 3.78 & 3.93 & 4.40 & 4.38 & 3.73 \\
    
    CycleGAN & 3.20 & 3.93 & 4.38 & 4.35 & 3.45\\
    
    UDDN & 3.53 & 4.05 & 4.33 & 4.28 & 3.63 \\
    
    DR2 & \underline{4.43} & \underline{4.13} & \underline{4.55} & \underline{4.48} & \underline{4.15} \\
    
    GDP & 4.38 & 4.03 & 4.33 & 4.30 & 3.98 \\
    
    \cellcolor[HTML]{E8E5E5} PFAD & \cellcolor[HTML]{E8E5E5}\textbf{4.78} & \cellcolor[HTML]{E8E5E5}\textbf{4.38} & \cellcolor[HTML]{E8E5E5}\textbf{4.73} & \cellcolor[HTML]{E8E5E5}\textbf{4.78} & \cellcolor[HTML]{E8E5E5}\textbf{4.63}\\
    \hline
    \end{tabular}
\caption{Evaluated by radiologists using Likert 5-scale scoring system. Higher scores indicate higher performance. Only the mean results are shown here, complete results can be found in Table~\ref{table_all_clinical} in supplementary material B.2.}
\label{table_clinical}
\end{small}
\end{table}

Next, we validate our method on real clinical T2-weighted images. Figure~\ref{clinical}(a) shows the results. For the supervised approach, Pix2pix causes distortion of normal tissue structures, such as intrahepatic bile ducts and branches of the hepatic veins as shown in Figure 7(b). For the unsupervised approaches, CycleGAN fails to remove motion artifacts and even makes them worse (Figure 7(c)). Similarly, UDDN is able to deal with weak motion artifacts, but not severe artifacts as shown in Figure 7(d). For the diffusion-based approaches, DR2 causes the image to be smooth when removing weak artifacts and fails to deal with severe artifacts (Figure~\ref{clinical}(e)). GDP produces the artifact-like tissues when removing weak artifacts, and also fails to deal with severe artifacts (Figure~\ref{clinical}(f)). In contrast, our method exhibited superior performance in effectively artifact removal while maintaining more of the correct tissue textures as shown in Figure~\ref{clinical}(g). It outperforms other approaches in facing clinical real artifacts.

Table~\ref{table_clinical} shows the clinical evaluation results. All results underwent Mann-Whitney U test, and most of the metrics have statistical significance. Our method shows superior performance and achieves the highest scores across all evaluation metrics. Additionally, we observe that diffusion model-based approaches generally outperforms GANs-based approaches clinically. This is because the GANs-based approaches are limited to the transformation between their finite distributions. Therefore the diffusion model-based approaches have high application value in clinical diagnosis.

\subsection{Ablation Study}

In this section, we validate the effectiveness of the pixel-frequency domain and the alternate complementary masks. All experiments are conducted on HCP dataset.

\subsubsection{Validation of Pixel-Frequency Domain}
We conduct ablation experiment in pixel domain and frequency domain to verify the important roles they play in removing artifacts.

From Table~\ref{dual_domain}, it can be seen that artifact removal in frequency domain plays an important role in maintaining correct tissue textures and affects most of metrics such as PSNR and SSIM. It is difficult to maintain the correct tissue textures by artifact removal only in the pixel domain. Meanwhile, pixel domain mainly plays an important role in visual sharpness metrics such as Lpips. Although the difference between the evaluation results in frequency domain only and pixel-frequency domain is small, it can be seen from Figure~\ref{show_compare} in supplementary material B.3 that the image quality recovered in pixel-frequency domain is better.

\subsubsection{Validation of Masks and Weight}
In this section we will verify the effectiveness of $\mathcal{M}$ and the influence of weight $\omega$ in two domains respectively.

From Table~\ref{table:4}, it's evident that $\mathcal{M}$ and $\omega$ plays a crucial role in artifact removal outcomes. Without $\mathcal{M}$, the diffusion model only relies on low-frequency image information, resulting in high-frequency information recovery that doesn't match the original image. When $\mathcal{M}$ is used, failure to use $\omega$ in either the pixel or frequency domain leads to excessive artifact information in the guidance information, resulting in poor image recovery quality.

\subsection{Hyperparameter Study}

In order to explore the impact of hyperparameters on experiments, we study the parameter $a$ in (\ref{eq:16}). Table~\ref{hyperparameter} shows the results under different values of $a$.

As can be seen from Table~\ref{hyperparameter}, when $a$ increases to 0.9, the PSNR and SSIM metrics decrease significantly, while the Lpips metric is relatively better. This shows that the guidance of frequency domain information focuses more on maintaining correct tissue textures, while the pixel domain information guides the model to generate clearer images. We use the total metric as a reference, combining higher-benefit metrics through addition and lower-benefit metrics through subtraction. In our experiments, we choose the value of $a$ for the best case of the total metric, where $a$ is equal to 0.7.

\section{Conclusion}
\begin{table}[t]
\begin{small}
\setlength{\tabcolsep}{2.4pt}
\centering
    \begin{tabular}
    {c|c|ccccccc}
    \hline
    Freq. & Pixel & PSNR & SSIM & Lpips & FSIM & VIF & VSI & GMSD \\
    \hline
    $\scalebox{0.75}{\usym{2613}}$ & $\checkmark$ & 23.89 & 0.590 & 0.190 & 0.830 & 0.280 & 0.951 & 0.126\\
    
    $\checkmark$ & $\scalebox{0.75}{\usym{2613}}$ & 27.60 & 0.829 & 0.154 & 0.896 & 0.423 & 0.970 & 0.063\\
    
    $\checkmark$ & $\checkmark$ & \textbf{27.60} & \textbf{0.829} & \textbf{0.147} & \textbf{0.896} & \textbf{0.426} & \textbf{0.970} & \textbf{0.061}\\
    \hline
    \end{tabular}

\caption{Ablation study of pixel-frequency domain.}
\label{dual_domain}
\end{small}
\end{table}

\begin{table}[!t]
\begin{small}
\setlength{\tabcolsep}{1.9pt}
\centering
    \begin{tabular}
    {c|cc|ccccccc}
    \hline
    \multirow{2}{*}{$\mathcal{M}$} & \multicolumn{2}{c|}{$\omega$} & \multirow{2}{*}{PSNR} & \multirow{2}{*}{SSIM} & \multirow{2}{*}{Lpips} & \multirow{2}{*}{FSIM} & \multirow{2}{*}{VIF} & \multirow{2}{*}{VSI} & \multirow{2}{*}{GMSD}\\
    \cline{2-3}
     & Freq. & Pixel & & & & \\
    \hline
    $\scalebox{0.75}{\usym{2613}}$ & $\scalebox{0.75}{\usym{2613}}$ & $\scalebox{0.75}{\usym{2613}}$ & 26.63 & 0.815 & 0.150 & 0.889& 0.402 & 0.996 & 0.067\\
    
    $\checkmark$ & $\scalebox{0.75}{\usym{2613}}$ & $\scalebox{0.75}{\usym{2613}}$ & 26.01 & 0.688 & 0.342 & 0.814 & 0.297 & 0.946 & 0.113\\
    
    $\checkmark$ & $\checkmark$ & $\scalebox{0.75}{\usym{2613}}$ & 26.08 & 0.696 & 0.335 & 0.817 & 0.302 & 0.947 & 0.112\\
    
    $\checkmark$ & $\scalebox{0.75}{\usym{2613}}$ & $\checkmark$ & 26.42 & 0.730 & 0.295 & 0.833 & 0.328 & 0.952 & 0.104\\
    
    $\checkmark$ & $\checkmark$ & $\checkmark$ & \textbf{27.60} & \textbf{0.829} & \textbf{0.147} & \textbf{0.896} & \textbf{0.426} & \textbf{0.970} & \textbf{0.061}\\
    \hline
    \end{tabular}
\caption{Ablation study of the masks ($\mathcal{M}$) and weight ($\omega$).}
\label{table:4}
\end{small}
\end{table}

\begin{table}[!t]
\begin{small}
\setlength{\tabcolsep}{2.8pt}
\centering
\begin{tabular}
    {c|cccccccc}
    \hline
    $a$ & PSNR & SSIM & Lpips & FSIM &VIF & VSI & GMSD & Total\\
    \hline
    0.1 & 27.60 & 0.829 & 0.154 & 0.896 & 0.423 & 0.970 & 0.063 & 30.501\\
    
    0.3 & 27.60 & 0.829 & 0.153 & 0.896 & 0.424 & 0.970 & 0.062 & 30.504\\
    
    0.5 & 27.60 & 0.830 & 0.151 & 0.897 & 0.425 & 0.970 & 0.062 & 30.509\\

    0.7 & 27.60 & 0.829 & 0.147 & 0.896 & 0.426 & 0.970 & 0.061 & \textbf{30.513}\\

    0.9 & 27.56 & 0.817 & 0.135 & 0.894 & 0.425 & 0.968 & 0.060 & 30.469\\
    \hline
    \end{tabular}
\caption{Results under different values of $a$.}
\label{hyperparameter}
\end{small}
\end{table}

In this paper, we propose a novel diffusion model inference architecture for motion artifact removal. By introducing the complementary masks in pixel-frequency domain, we use partial image information as guidance to reconstruct clean images with the diffusion model's excellent generative capability. Experimental results demonstrate our method's effectiveness in removing motion artifacts and exhibiting excellent performance on real clinical images. However, in simulated experiments, our method performs sub-optimally on some evaluation metrics compared to supervised methods. This may be due to their ability to adapt well to the distribution of simulated images, which also results in a significant drop in their performance when dealing with real artifacts. In summary, we propose a novel artifact removal method for medical image analysis, aiming to mitigate the impact of artifacts on clinical diagnosis. The proposed method may contribute to radiologist making more correct diagnoses.

\section{Acknowledgments}
This work was supported in part by the National Key R\&D Program of China under Grant No.2023YFA1008600, in part by the National Natural Science Foundation of China under Grants U22A2096 and 62441601, in part by the Shaanxi Province Core Technology Research and Development Project under Grant 2024QY2-GJHX-11, in part by the Fundamental Research Funds for the Central Universities under Grant QTZX23042, in part by the National Science Foundation for Young Scientists of China under Grant 82302130.

\bibliography{aaai25}
\clearpage
\appendix
\onecolumn
\begin{center}
{\bf {\LARGE Supplementary Material}} \\
\vspace{0.05in}
\end{center}

\section{A. Experimental Details}
\subsection{A.1 Datasets}\label{dataset}
\subsubsection{Brain} HCP dataset is obtained by a Siemens 3T system with a 3D spin-echo sequence. The imaging parameters for acquiring the HCP data are as follows: TR = 3200 ms, TE = 565 ms,  voxel size = 0.7 mm $\times $ 0.7 mm $\times $ 0.7 mm, matrix size = 320 $\times $ 320, number of slices = 256, echo train duration = 1105, echo spacing = 3.53 ms, and the phase encoding direction = anterior-posterior. We selected 3000 motion-free images from 150 subjects with 20 images per subject for training the diffusion model. Additionally, 800 images were chosen from 40 subjects with 20 images per subject for generating motion-corrupted images and evaluation. All images are center-cropped to the size of 256 $\times$ 256 to remove redundant areas that do not contain brain information.

\subsubsection{Knee} 
We use the knee MRI data from fastMRI dataset. The number of coils of this data is 15. The acquisition parameters are as follows: TR = 2200 - 3000 ms, TE = 27 - 34 ms, matrix size = 320 $\times$ 320, in-plane resolution = 0.5 mm $\times$ 0.5 mm, slice thickness = 3 mm, and phase encoding direction = anterior-posterior. We selected 3000 motion-free images for training and 800 images for generating motion-corrupted images and evaluation.

\subsubsection{Abdominal} 
This dataset is obtained by a 3T Philips Ingenia MR system with the following scan parameters: TR = 2130.6 ms, TE = 80 ms, slice thickness/intersection gap = 5/0 mm, echo time = 80 ms, field of view = 100 mm $\times$ 100 mm, acquisition matrix = 288 $\times$ 288, the phase encoding direction = anterior-posterior. After acquisition, zero padding was applied to the acquired \textit{k}-space, resulting in image sizes of 640 $\times$ 640 after padding. All images are center-cropped to the size of 512 $\times$ 512. We selected 3002 motion-free images from 94 subjects for training and 576 images from 18 subjects to generate motion-corrupted images and evaluate the performance of our method. Additionally, 40 real motion-corrupted images from clinical settings are used for qualitative and radiologist evaluations. 

\subsection{A.2 Simulation of Motion Artifacts}
Similar to previous works \cite{tamada2020motion, oh2021unpaired}, the simulation of motion artifacts can be expressed as follows,
\begin{equation}
\label{eq:18}
\hat{x}(k_x,k_y)= 
\begin{aligned}
\begin{cases}
    x(k_x,k_y)e^{-j\Phi(k_y)}, &k_y \in \mathbb{K}\\
    x(k_x,k_y), &\text{otherwise},
\end{cases}
\end{aligned}
\end{equation}
where $\hat{x}$ and $x$ denotes refer to the
motion-corrupted and motion-free \textit{k}-space data, $j=\sqrt{-1}$, $k_x$ and $k_y$ are the indices along the read-out and phase encoding directions, respectively. $\Phi(k_y)$ is the displacement in radian along the phase encoding direction, and $\mathbb{K}$ is the position in the phase-encoding direction where the perturbations occur.

For brain and knee datasets, we simulate random motion artifacts of translation and rotation. $\Phi(k_y)$ in (\ref{eq:18}) can be represented as follows,
\begin{equation}
\label{eq:19}
\Phi(k_y)=
\begin{aligned}
\begin{cases}
\mathcal{R}\{k_y\Delta_k\}, & |k_y|>k_0\\
0, & \text{otherwise},
\end{cases}
\end{aligned}
\end{equation}
where $\Delta_k$ denotes the degree of motion along $y$ direction, $\mathcal{R}$ denotes the rotation angle of the image in pixel domain, and $k_0$ is the delay time of the phase perturbation \cite{tamada2020motion}. In this simulation, $k_0$ is fixed to $\pi/10$, $\Delta_k$ is adjusted to be 2.5--3.0 cm, and $\mathcal{R}$ is sampled from $[-2^\circ, 2^\circ]$.

For abdominal dataset, we simulate respiratory 
motion artifacts as follows,
\begin{equation}
\label{eq:20}
\Phi(k_y)=
\begin{aligned}
\begin{cases}
k_y\Delta_k \sin(mk_y+n), & |k_y|>k_0,\\
0, & \text{otherwise},
\end{cases}
\end{aligned}
\end{equation}
where $\Delta_k$, $m$, and $n$ denote the amplitude, period, and phase perturbation, respectively. In this simulation, $\Delta_k$ is adjusted to be 1.1--1.2 cm, $k_0$ is fixed to $\pi/10$, $m$ is randomly sampled from $[0.1, 5.0]$, and $n$ is randomly sampled from $[0, \pi/4]$.

\subsection{A.3 Training and Evaluation Details}
\subsubsection{Training}
We use guided diffusion\cite{dhariwal2021diffusion} for training. We set the number of the channel dimensions to 128, the number of residual block to 2, the resolution of self-attention residual blocks to ${32, 16, 8}$, and the number of channels per attention head to 64. The channel multipliers are set to $\{1,1,2,2,4,4\}$,  $\{1,1,2,4,4,4\}$ and $\{0.5,1,1,2,2,4,4\}$ for size 256, 320, and 512, respectively. We set the batch size to 4, the total number of timesteps to 1000, and learning rate to $1.0 \times 10^{-4}$. We trained the model with 50k, 200k, and 500k iterations respectively on size 256, 320, and 512. The diffusion model training is conducted on two RTX 4090 GPUs.

\subsubsection{Evaluation}
We use five metrics to quantitatively evaluate the results on the simulated images: peak signal-to-noise ratio (PSNR), structural similarity index measure (SSIM) \cite{wang2004image}, learned perceptual image patch similarity (Lpips) \cite{zhang2018unreasonable}, feature similarity index measure (FSIM) \cite{zhang2011fsim}, visual information fidelity (VIF) \cite{sheikh2006image}, visual saliency-induced Index (VSI) \cite{zhang2014vsi} and gradient magnitude similarity deviation (GMSD) \cite{xue2013gradient}. Meanwhile, considering that the real clean images paired to motion-corrupted images are hard to obtain, we perform a qualitative evaluation on the clinical images. Two young radiologists with 3 years of diagnostic experience independently evaluated the denoising results without knowing any clinical details or biopsy results, using a Likert 5-scale \cite{likert1932technique} scoring system where higher scores indicate better performance. Images were shown in a random order to avoid bias. The evaluation metrics included artifact removal performance, image blurriness, anatomical deformation, diseased tissue deformation, and overall image quality. If the opinions of two young radiologists were consistent, their scores were used as the final scores. Otherwise, another radiologist with 20 years of experience conducted a further review and made final decisions.

\section{B. Additional Experimental Results}
\subsection{B.1 Comparison of Visualization Results on Simulated Images}
\begin{figure*}[h]
    \centering
    \includegraphics[width= 6.4in]{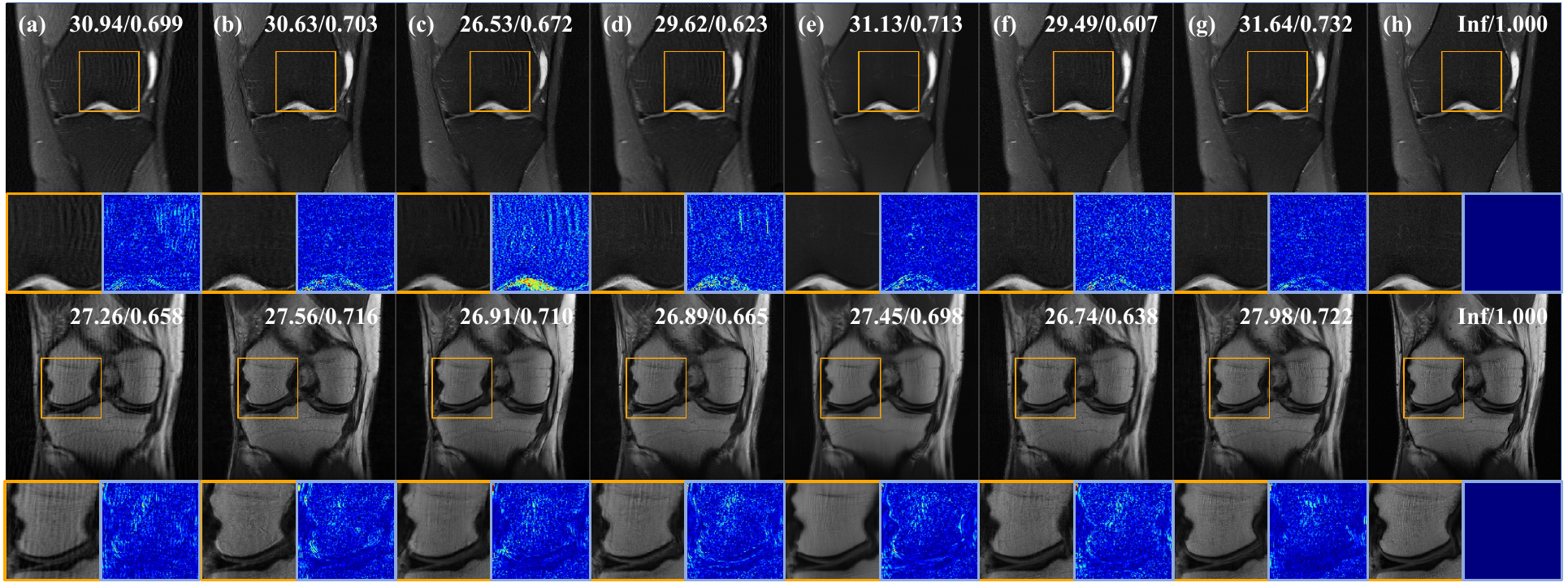}
    \caption{Comparison of results on simulated knee images: (a) Artifact images, (b) Pix2pix, (c) CycleGAN, (d) UDDN, (e) DR2, (f) GDP, (g) PFAD (ours), and (h) the ground-truth. PSNR and SSIM of each image are shown in the corner of images.}
    \label{show_knee}
\end{figure*}

\begin{figure*}[h]
    \centering
    \includegraphics[width= 6.4in]{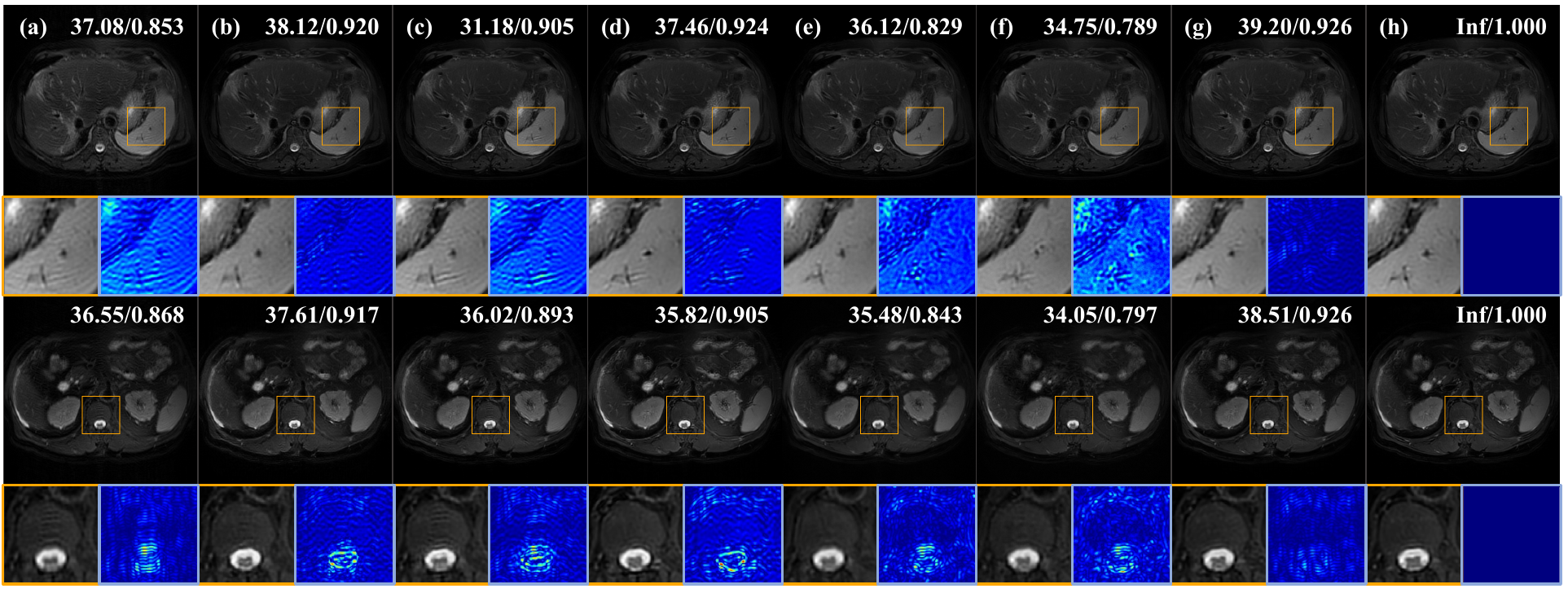}
    \caption{Comparison of results on simulated  abdominal images: (a) Artifact images, (b) Pix2pix, (c) CycleGAN, (d) UDDN, (e) DR2, (f) GDP, (g) PFAD (ours), and (h) the ground-truth. PSNR and SSIM of each image are shown in the corner of images.}
    \label{show_abdominal}
\end{figure*}

We performed visualization results on the knee images and the abdominal images, as shown in Figure~\ref{show_knee} and Figure~\ref{show_abdominal}, respectively. From the results, it can be seen that our method achieves the best in texture detail retention, image naturalness and clarity.

\subsection{B.2 Complete Results of Qualitative Evaluation}
Our qualitative evaluation results of radiologists are shown in the form of mean $\pm$ variance in Table~\ref{table_all_clinical}. It can be seen that our method achieves the best performance in all metrics, and its clinical performance is relatively stable due to its small variance.
\begin{table}[h]
\setlength{\tabcolsep}{6.0pt}
\renewcommand\arraystretch{1.15}
\centering
   \begin{tabular}
    {c|ccccc}
    \hline
    Methods & Artifacts & Blurring & Anatomical & Diseased & Overall\\
    \hline
    Pix2pix & 3.78$\pm$0.95 & 3.93$\pm$0.80 & 4.40$\pm$0.67 & 4.38$\pm$0.70 & 3.73$\pm$0.82 \\
    
    CycleGAN & 3.20$\pm$0.91 & 3.93$\pm$0.73 & 4.38$\pm$0.67 & 4.35$\pm$0.66 & 3.45$\pm$0.71\\
    
    UDDN & 3.53$\pm$1.18 & 4.05$\pm$0.71 & 4.33$\pm$0.69 & 4.28$\pm$0.72 & 3.63$\pm$0.93 \\
    
    DR2 & \underline{4.43}$\pm$0.75 & \underline{4.13}$\pm$0.72 & \underline{4.55}$\pm$0.68 & \underline{4.48}$\pm$0.68 & \underline{4.15}$\pm$0.70 \\
    
    GDP & 4.38$\pm$0.74 & 4.03$\pm$0.80 & 4.33$\pm$0.76 & 4.30$\pm$0.85 & 3.98$\pm$0.73 \\
    
    \cellcolor[HTML]{E8E5E5} PFAD & \cellcolor[HTML]{E8E5E5}\textbf{4.78}$\pm$0.48 & \cellcolor[HTML]{E8E5E5}\textbf{4.38}$\pm$0.74 & \cellcolor[HTML]{E8E5E5}\textbf{4.73}$\pm$0.55 & \cellcolor[HTML]{E8E5E5}\textbf{4.78}$\pm$0.53 & \cellcolor[HTML]{E8E5E5}\textbf{4.63}$\pm$0.54\\
    \hline
    \end{tabular}
\caption{Denoising results evaluated by radiologists using Likert 5-scale scoring system (mean$\pm$standard deviation). Higher scores indicate higher performance.}
\label{table_all_clinical}
\end{table}

\subsection{B.3 Comparison of Visualization Results on Ablation Study of Pixel-Frequency Domain}
\begin{figure*}[h]
    \centering
    \includegraphics[width= 4.2in]{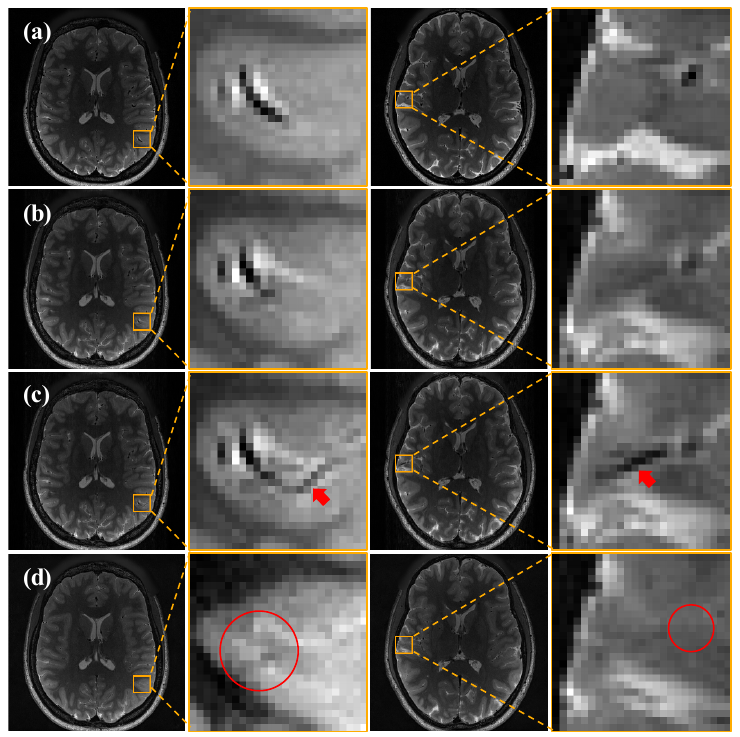}
    \caption{Comparison of visualization results on ablation study of pixel-frequency domain: From top to bottom, each row represents: (a) the ground-truth, (b) PFAD, (c) Frequency domain only, (d) Pixel domain only. The yellow box indicates the zoomed-in visualization area, the red arrow points to the area where the corrugated structure appears, and the red circle indicates the area where the texture disappears.}
    \label{show_compare}
\end{figure*}

Figure~\ref{show_compare} shows that performing artifact removal in pixel-frequency domain is the best. As shown in Figure~\ref{show_compare}(c) and Figure~\ref{show_compare}(d), only performing artifact removal in the frequency domain will make the recovered image unnatural and have a ripple structure, while only performing artifact removal in the pixel domain cannot maintain texture information, causing some textures to disappear.

\subsection{B.4 Experiment on Cutoff}
We perform the following experiment on the cutoff frequency of the filter.
\begin{table}[h]
\setlength{\tabcolsep}{2.4pt}
\centering
    \begin{tabular}
    {c|ccccccc}
    \hline
    Cutoff & PSNR & SSIM & Lpips & FSIM & VIF & VSI & GMSD \\
    \hline
    $\pi/5$ & 25.89 & 0.711 & 0.283 & 0.826 & 0.306 & 0.950 & 0.112\\
    
    $\pi/10$ & 27.60 & 0.829 & 0.147 & 0.896 & 0.426 & 0.970 & 0.061\\
    
    $\pi/20$ & 26.34 & 0.779 & 0.180 & 0.867 & 0.343 & 0.962 & 0.085\\
    \hline
    \end{tabular}

\caption{Experiment on the cutoff frequency of the filter.}
\label{cutoff}
\end{table}

If the cutoff is too small, the useful guided information is too small to recover the correct texture, if the cutoff is too large, the guided information contains too many artifact information, both of them lead to bad results. Therefore, it is necessary to correctly select the cutoff frequency in clinical diagnosis.

\subsection{B.5 Experiment on Grid Size}
We have used a checkerboard mask with the grid size of 16. We perform an experiment on the grid size. The results are as follows.
\begin{table}[h]
\setlength{\tabcolsep}{2.4pt}
\centering
    \begin{tabular}
    {c|ccccccc}
    \hline
    Size & PSNR & SSIM & Lpips & FSIM & VIF & VSI & GMSD \\
    \hline
    1 & 27.57 & 0.829 & 0.153 & 0.896 & 0.423 & 0.970 & 0.062\\

    4 & 27.57 & 0.829 & 0.153 & 0.896 & 0.423 & 0.970 & 0.063\\

    16 & 27.60 & 0.829 & 0.147 & 0.896 & 0.426 & 0.970 & 0.061\\
    
    64 & 27.42 & 0.827 & 0.154 & 0.895 & 0.420 & 0.969 & 0.064\\
    
    128 & 27.24 & 0.824 & 0.155 & 0.893 & 0.416 & 0.969 & 0.065\\
    \hline
    \end{tabular}

\caption{Experiment on the grid size of checkerboard mask.}
\label{grid_size}
\end{table}

Large grid size destroys the distribution of useful information and small grid size does not significantly destroy the distribution of artifact, both of them lead to poorer results, the former is worse.
\end{document}